\documentstyle [aps,12pt,epsf]{revtex} \bibstyle{unsrt}
  \large \normalsize

\begin{document}

\begin{center}
\vspace*{1.0cm}

{\Large\bf Optimal Control of Molecular Motion Expressed Through Quantum Fluid Dynamics}

\vspace{0.5cm}

Bijoy K Dey$^{*}$, Herschel Rabitz 

\vspace*{0.2cm}

{\small \sl Department of Chemistry, Princeton University, Princeton, New Jersey}\\
\vspace{0.2cm}

and

Attila Askar

\vspace*{0.2cm}

{\small \sl Department of Mathematics, Koc University, Istanbul, Turkey}\\

\vspace*{1.2cm}
\end{center}


\vspace{4.6cm}
\footnotemark[1]{{\bf Present address :} Department of Chemistry, Chemical Physics Theory Group,\\ \hspace*{4.0cm}University of Toronto, Toronto, Canada}
\parindent=0.5cm
\newpage

\begin{center}

\vspace*{0.7cm}

{\bf Abstract}
\end{center}

~~~~~~~~~~~~A quantum fluid dynamic(QFD) control formulation is presented for optimally manipulating atomic and molecular systems. In QFD the control quantum system is expressed in terms of the probability density $\rho $ and the quantum current j. This choice of variables is motivated by the generally expected slowly varying spatial-temporal dependence of the fluid dynamical variables. The QFD approach is illustrated for manipulation of the ground electronic state dynamics of HCl induced by an external electric field. 

\section{Introduction}
Manipulating the outcome of quantum dynamics phenomena by a properly tailored external control is a topic of increasing activity [1-15]. Problems where the external control is electromagnetic have received the most attention, although other applications arise as well. Various implementations of quantum control have been experimentally realized [16-25]. A variety of control strategies have been suggested, and optimal control theory(OCT) provides the most general framework for acheiving field designs. Such designs will generally require further refinement in the laboratory through learning techniques to overcome design uncertainities [17-20]. A basic difficulty in attaining the control designs is the computational effort called for in solving the time-dependent Schroedinger equation, often repeatedly in an iterative fashion. This paper introduces the quantum fluid dynamic(QFD) control formulation to simplify this task. Calculations have shown [28, 29] that QFD is capable of being much more efficient than conventional methods(e.g., FFT propagation), and this savings should carry over to the control design task. This paper will show how OCT can be combined with QFD. 

The theoretical basis for dynamic control [2-13] is to create non-stationary states of one's choice, by optimally designing the control field. Typically, the problem is posed as seeking an optimal field to drive a quantum wave packet to a desired target at a chosen time t=T. In the traditional approach [3-10] to quantum optimal control an objective design functional $\bar{J}$ is defined, which depends on the system wave function, a wave function like Lagrange multiplier, and the external field. Minimization of the objective functional leads to the identification of external field(s) capable of delivering a specific outcome. This process requires solving for the complex oscillatory wave function and the similarly behaved Lagrange multiplier, and due care is needed for their proper representation in a suitable basis for capturing their behaviour. Often the rapidly varying spatio-temporal behaviour of these functions necesitates the use of many unknowns in the basis.

This paper explores an alternative formulation for OCT to design the electric field. The formulation is based on the fluid dynamic view point of quantum mechanics [26-29], which by-passes the typically oscillatory  nature of the wave function to exploit the generally smooth behaviour of the real density and the quantum current variables. Recent illustrations have demonstrated the smooth spatial and temporal nature of the variables and the ability to discretize them on a relatively small number of grid points [28, 29]. As background in section 2 we give a brief summary of the QFD formulation of quantum mechanics. Section 3 presents OCT within the framework of QFD for designing an electric field to meet a specific objective. Section 4 applies the OCT-QFD formulation for the manipulation of HCl. Section 5 concludes the paper.

\section{Quantum Fluid Dynamics}
~~~~~~~~~~~~~The treatment below considers a single particle of reduced mass m, but the QFD formulation has an immediate extension to many particles. The time-dependent Schr\"{o}dinger equation is given by
\begin{eqnarray}
[-\frac{\hbar ^2}{2m}\nabla ^2 + V+ V_{ext}]\Psi ({\bf x},t)=i\hbar \frac{\partial}{\partial t}\Psi ({\bf x},t)
\end{eqnarray}
where V typically confines the particle in a locale and $V_{ext}$ is the control taken here as $-\mu ({\bf x})\cdot E(t)$ with E(t) being the electric field and $\mu ({\bf x})$ the dipole moment. Substituting $\Psi ({\bf x},t)=A({\bf x},t)e^{iS({\bf x},t)/\hbar}$, where A and S are real functions, into Eq.(1) and separating the real and imaginary parts one easily obtains two equations. The imaginary part yields the continuity equation
\begin{eqnarray}
\frac{\partial \rho}{\partial t}+\nabla \cdot (\rho {\bf v})=0
\end{eqnarray}
and the real part the following equation for the phase S
\begin{eqnarray}
\frac{\partial S}{\partial t}+\frac{\nabla S\cdot \nabla S}{2m}+V_{eff}=0
\end{eqnarray}
where $V_{eff}=V+V_{ext}+V_q$ with $V_q=-\frac{\hbar ^2}{2m}\frac{\nabla ^2\rho ^{1/2}}{\rho ^{1/2}} = -\frac{\hbar ^2}{2m}[\nabla ^2ln\rho ^{1/2}+(\nabla ln\rho ^{1/2})^2]$, $\rho =|\Psi |^2$ and {\bf v}=$\frac{\nabla S}{m}$.

Equation (3) has the form of the classical Hamilton-Jacobi equation with an extra 'quantum potential' term $V_q$. This equation can be transformed into one for the evolution of the velocity vector {\bf v} by taking the gradient to give
\begin{eqnarray}
\frac{\partial}{\partial t}{\bf v}=-({\bf v}\cdot \nabla ){\bf v}-\frac{1}{m}\nabla (V_{eff})
\end{eqnarray}
Defining the quantum current as
\begin{eqnarray*}
{\bf j}({\bf x},t)=-\frac{\hbar}{m}Im[\Psi ^*({\bf x},t)\nabla \Psi ({\bf x},t)]=\rho ({\bf x},t){\bf v}({\bf x},t),
\end{eqnarray*}
one readily obtains the equation of motion for {\bf j} by substitution of $\frac{\partial \rho}{\partial t}$ and $\frac{\partial}{\partial t}${\bf v} from Eqs.(2) and (4) as
\begin{eqnarray}
\frac{\partial }{\partial t}{\bf j}=-{\bf v}(\nabla \cdot {\bf j})-({\bf j}\cdot \nabla ){\bf v}-\frac{\rho}{m}\nabla V_{eff}
\end{eqnarray}
Eqs.(2) and (3) or (2) and (5) describe the motion of a quantum particle within the QFD formulation of quantum mechanics. The motion of a quantum particle is governed by the current vector {\bf j} and the density $\rho $ in Eqs.(2) and (5). Although the QFD equations resemble those of classical fluid dynamics, their quantum identity prevails due to the presence of the  potential $V_q$ which has no classical analogue. Equivalently, the QFD equations may be viewed as those of a ``classical'' fluid with a highly non-linear constitutive law prescribed by $V_{eff}$. Various Eulerian or Lagrangian means can be exploited to solve the QFD equations [28,29], and available fluid dynamics codes may be adopted to treat these equations [30]. The essential simplifying feature of the QFD equations is that $\rho $ and ${\bf j}$ or $\rho $ and S are often slowly varying, which is evident from quantum dynamics calculations [28,29], thereby permitting relatively coarse gridding. Despite the non-linear nature of the QFD equations, the general smoothness of $\rho$ and ${\bf j}$ observed lead to significant computational savings [28,29].

\section{Control Expressed Within Quantum Fluid\\ Dynamics}
Quantum OCT seeks the design of an external field to fulfill a particular dynamical objective. This section will provide the working equations for OCT-QFD to design an optimal electric field that drives a quantum wave packet to a desired objective at the target time t=T. The OCT-QFD formulation could be expressed in the usual way in terms of the Schr\"{o}dinger equation where QFD would only act as a solution procedure. Here we will present a general approach by writing OCT directly in terms of QFD. As an example the control of a non-rotating diatomic molecule will be used as a simple illustration of the concepts. The treatment of a fully general target expectation value $\Theta _T=<\Psi(T)|\Theta|\Psi(T)>$ may be considered with QFD, but here we will only treat the common case where the operator $\Theta (x)$ is only position x dependent. Then the goal is to steer $\Theta_{T}$

\begin{eqnarray}
\Theta_T=\int_{x_{l}}^{x_{r}}\Theta(x)\rho (x,T)dx
\end{eqnarray}
as close as possible to the desired value $\Theta ^d$.

The active spatial control interval is taken as $x_l \le x \le x_r$ over the time $0 \le t \le T$ that the control process occurs. We desire to minimize the cost functional $J_{cost}=J_{target}+J_{field}$ where $J_{target}$ and $J_{field}$ are given by 
\begin{eqnarray}
J_{target}=\frac{1}{2}\omega _x(\Theta _T-\Theta ^d)^2 \hspace*{1cm} and \hspace*{1cm} J_{field}=\frac{1}{2}\omega _e\int _{0}^{T}E^2(t)dt
\end{eqnarray}
with $\omega _e $ and $\omega _x$ being the positive weights balancing the significance of the two terms. The second term represents the penalty due to the fluence of the external field. The minimization of $J_{cost}$ with respect to E(t) must be subject to the satisfaction of the equations of motion for $\rho $ and ${\bf j}$ in Eqs.(2) and (5).  We may fulfill this constraint by introducing the unconstrained cost functional as 
\begin{eqnarray}
\bar{J}&=&J_{cost}-\int_0^T\int_{x_{l}}^{x_{r}}\lambda _1(x,t)[\frac{\partial \rho(x,t)}{\partial t}+\frac{\partial j(x,t)}{\partial x}]dx dt\\\nonumber
&&-\int_0^T\int_{x_{l}}^{x_{r}}\lambda _2(x,t)[\frac{\partial j(x,t)}{\partial t}+\frac{\partial }{\partial x}(\frac{j^2}{\rho })+\frac{\rho}{m}\frac{\partial }{\partial x}(V+V_q+V_{ext})]dx dt
\end{eqnarray}  
where $\lambda _1(x,t)$ and $\lambda _2(x,t)$ are Lagrange's multiplier functions. \\

An optimal solution satisfies $\delta \bar{J}=0$, which is assured by setting each of the functional derivatives with respect to $\lambda _1$, $\lambda _2$, $\rho $, {\bf j} and E to zero. The first two, i.e., the functional derivatives with respect to $\lambda _1$ and $\lambda _2$ regenerate the QFD equations in Eq.(2) and (5). The three others are obtained in the forms :

\begin{eqnarray}
\frac{\partial \lambda _2}{\partial t}+\frac{\partial}{\partial x}(\lambda _2v_{\lambda})+S_1[\rho ,j,\lambda _2]=0
\end{eqnarray}

\begin{eqnarray}
\frac{\partial \lambda _1}{\partial t}+\frac{\partial}{\partial x}(\lambda _1v_{\lambda})-\lambda _2\frac{\partial}{\partial x}(V+V_q(\lambda _2)+V_{ext})+S_2[\rho ,j,\lambda _2]=0
\end{eqnarray}
and
\begin{eqnarray}
\frac{\delta \bar{J}}{\delta E(t)}=\int_{x_{l}}^{x_{r}}\lambda _2(x,t)\rho (x,t)\frac{\partial }{\partial x}\mu (x)dx +\omega _eE(t)=0
\end{eqnarray}
where
\begin{eqnarray}
S_1=2\frac{j}{\rho }\frac{\partial \lambda _2}{\partial x}
\end{eqnarray}

\begin{eqnarray}
S_2&=&-\frac{\lambda _2}{m}\frac{\partial}{\partial x}(V_q(\rho)-V_q(\lambda _2))-\frac{j^2}{\rho ^2}\frac{\partial \lambda _2}{\partial x}\\\nonumber
&&-\frac{\hbar ^2}{4m^2\rho ^{1/2}}\frac{\partial ^2}{\partial x^2}[\frac{1}{\rho ^{1/2}}\frac{\partial}{\partial x}(\lambda _2\rho )]\\\nonumber
&&+\frac{\hbar ^2}{4m^2\rho ^{3/2}}\frac{\partial ^2}{\partial x^2}\rho ^{1/2}\frac{\partial}{\partial x}(\lambda _2\rho )
\end{eqnarray}
and
\begin{eqnarray}
V_q=-\frac{\hbar ^2}{2m}\frac{\nabla ^2\lambda _2 ^{1/2}}{\lambda _2 ^{1/2}} = -\frac{\hbar ^2}{2m}[\nabla ^2ln\lambda _2 ^{1/2}+(\nabla ln\lambda _2 ^{1/2})^2]
\end{eqnarray}

The corresponding final conditions are
\begin{eqnarray}
\omega _x[\Theta_T-\Theta ^d]\Theta (x)-\lambda _1(x,T)=0
\end{eqnarray}
and
\begin{eqnarray}
\lambda _2(x,T)=0
\end{eqnarray}

Several other constraint expressions can be obtained by using equivalent forms of the continuity and dynamical equations. The form presented above is used in the subsequent numerical calculations. An alternative form in multi-dimensions symmetric between the QFD and Lagrange multiplier functions is presented in the Appendix.

The equations (9) and (10) for $\lambda _2$ and $\lambda _1$ respectively ressemble that of $\rho $ and {\bf j} with the only difference being the extra source terms $S_1$ and $S_2$. The source terms depend on $\rho $ and {\bf j}. $v_{\lambda}$ in the above equations is the 'velocity' associated with the Lagrange's multiplier and is given as $v_{\lambda}=\frac{\lambda _1}{\lambda _2}$. There are now two different quantum potential terms, one of which is a function of $\rho (x,t)$ and the other is a function of $\lambda _2(x,t)$. In this formalism the evolution of $\lambda _1(x,t)$ takes place by $V_q(\lambda _2)$ as well as the difference of the two types of quantum potential. In obtaining the above equations we have standardly assumed no variation of either $\rho (x,0)$ or j(x,0). Thus, we start from the initial value of $\rho (x,0)$ and j(x,0) to solve Eqs.(2) and (5). Eqs.(9) and (10) can be solved for $\lambda _2(x,t)$ and $\lambda _1(x,t)$ by integrating backward from time T using  $\lambda _1(x,T)$ and $\lambda _2(x,T)$ given in Eqs.(15) and (16) respectively. The equations (2), (5), (9) and (10) are non-linear thereby calling for iteration to solve(cf., the algorithm in Section 4). Finally the desired control electric field is given from Eq.(11) as 
\begin{eqnarray}
E(t)=-\frac{1}{\omega _e}\int_{x_{l}}^{x_{r}}\lambda _2(x,t)\rho (x,t)\frac{\partial }{\partial x}\mu (x)dx
\end{eqnarray}
\section*{4  Application to HCl} 

The OCT-QFD formulation will be applied to manipulating the vibrational motion of HCl on the ground electronic state. The initial density $\rho (x,0)=|\Psi (x)|^2$ was obtained from solving for the vibrational state from the equation
\begin{eqnarray}
-\frac{\hbar ^2}{m }\frac{\partial ^2\Psi (x)}{\partial x^2}+V(x)\Psi (x)=E\Psi (x)
\end{eqnarray}
using the Fourier grid Hamiltonian method [31,32] where m is the reduced mass of the HCl molecule and $V(x)$ is the truncated polynomial presented by Olgilvie [33] 
\begin{eqnarray}
[V(x)=\left\{ \begin{array}{r@{\quad for\quad}l} a_1(2\frac{x-x_e}{x+x_e})^2[1+\sum_{i=2}^{9}a_i(2\frac{x-x_e}{x+x_e})^{i-1}]-b_1 & x < 4 \\ A[1-tanh(x-4)]^{3/2} & 4 \le x \le 6.5 \\ 0 & x \ge 6.5 \end{array} \right.  ]
\end{eqnarray}
where $x_e$=2.4086 a.u. is the equilibrium bond length of HCl. The parameters in a.u. entering the potential function are $a_1=0.961914$, $a_2=-1.362999$, $a_3=0.86675$, $a_4=-0.49804$, $a_5=0.1727$, $a_6=0.2687$, $a_7=-1.977$, $a_8=2.78$, $a_9=4.89$, $b_1=0.169695$ and $A=-4.85\times 10^{-2}$. Since $\Psi (x)$ is a stationary real function we have zero initial flux j(x,0)=0. The initial $\rho (x,0)$ is nearly a Gaussian packet centered around $x_e$. The dipole function for HCl is given by [34]
\begin{eqnarray}
\mu (x)=c_1[g(x)+c_2 g^2(x)+c_3 g^3(x)]
\end{eqnarray}
where $g(x)=1-tanh(\beta (x-x_d))$ and the parameters are $c_1=0.279$, $c_2=-0.905$, $c_3=1.029$, $\beta =0.687$ and $x_d=2.555$. The following steps were carried out for implementation of the present OCT-QFD algorithm :\\
\newcounter{fig}
\begin{list}{\bfseries\upshape Step \arabic{fig}:}
{\usecounter{fig}
\setlength{\labelwidth}{2cm}\setlength{\leftmargin}{2.6cm}
\setlength{\labelsep}{0.5cm}\setlength{\rightmargin}{1cm}
\setlength{\parsep}{0.5ex plus0.2ex minus 0.1ex}
\setlength{\itemsep}{0ex plus0.2ex} \slshape}
\item Make an initial guess for the electric field E(t), which was zero in the present calculations.
\item Solve the coupled equations, viz., Eq.(2) and (5) for $\rho (x,t)$ and j(x,t) respectively starting from $\rho (x,0)$ and j(x,0). The solution was achieved here by using the Flux-corrected transport(FCT) algorithm [35] modified for the purpose of solving the QFD equations [28]. In doing so, we adopt the Eulerian numerical scheme.
\item Evaluate the final value for $\lambda _1(x,T)$ given by Eq.(15) and set $\lambda _2(x,T)$=0 by Eq.(16).
\item Solve Eqs.(9) and (10) for $\lambda _2(x,t)$ and $\lambda _1(x,t)$, respectively, by backward propagation using the same method as in step 2. Equations (9) and (10) have source terms which depend on $\rho (x,t)$ and j(x,t) calculated from step 2.
\item Calculate the difference between the left and right sides of Eq.(16) for use in the conjugate gradient method [36] and calculate $J_{cost}$ from Eq.(7).
\item Iterate steps 2 to step 6 until acceptable convergence is met.
\end{list}

The spatial range of the calculation was $0\le x \le 12$ a.u., and the time interval was $0\le t \le T$ with T=2000 a.u. The total number of spatial mesh points is 64 which gives $\Delta x=0.1875$ a.u. Similarly, the total number of time steps was 2048, which corresponds to $\Delta t=0.9765$ a.u. No special effort was made to optimize the grid points, as the purpose here is to demonstrate the QFD-OCT formulation. The weight $\omega _e$ in Eq.(7) was taken as $\frac{1}{2}$, and $\omega _x$ = $1000$. The target operator was $\Theta =x$ and $\Theta ^d=3.0 a.u.$.

\begin{figure}[!h]
\epsfxsize=6.0in
\hspace*{0.0cm}\epsffile{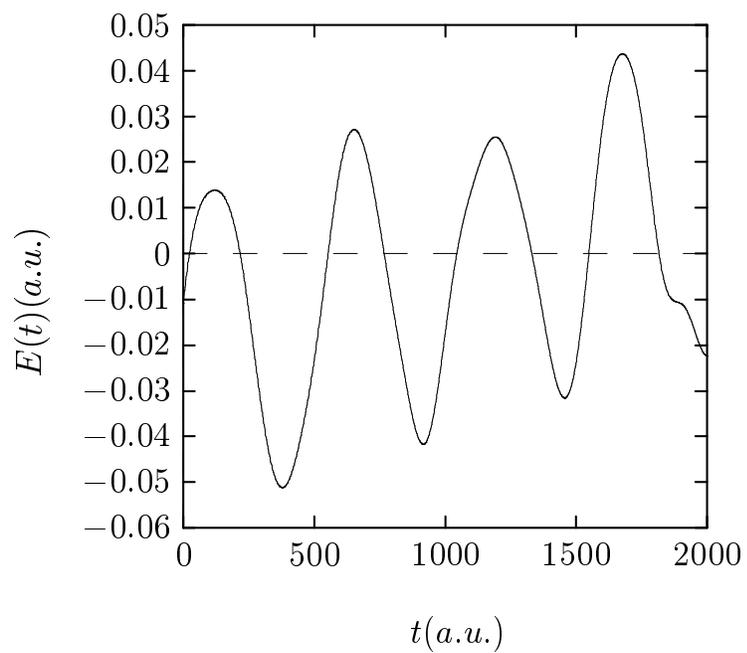}
\vspace*{-7.0cm}\caption{The optimal control field in a.u. shown as a function of time.}
\label{fig1}
\end{figure}

\begin{figure}[!h]
\epsfxsize=6.0in
\hspace*{0.0cm}\epsffile{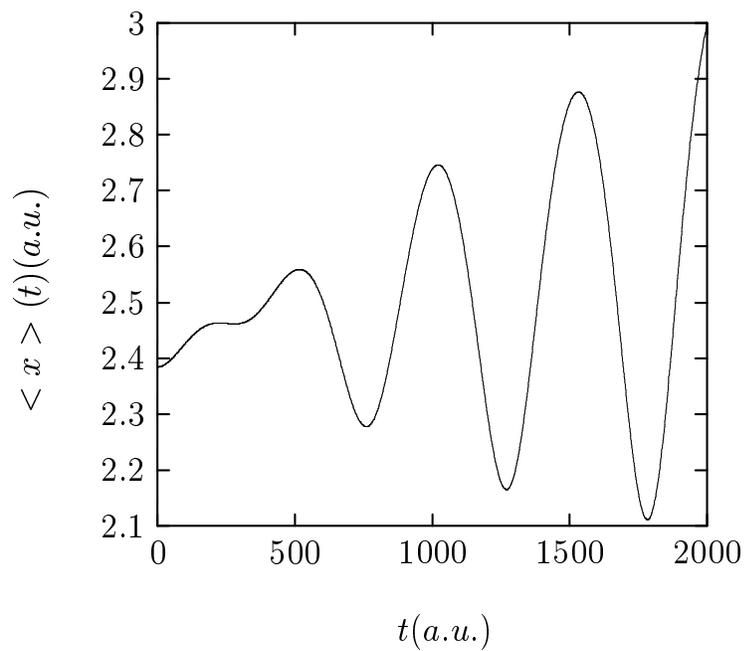}
\vspace*{-7.0cm}\caption{The expectation value $<x>$ shown as a function of time in a.u. The target value is $<x>$=3.0 a.u. at T=2000 a.u.}
\label{fig2}
\end{figure}

Figure \ref{fig1} shows the control field in atomic units. The slightly non-zero values of the field at the beginning and end could be arrested by placing additional costs if desired. This pulse excites several vibrational states(not shown here) mainly by a sequence of single quantum transitions. Figure \ref{fig2} shows the average distance $<x>$ as a function of time. The desired control value of $<x>$=3.0 a.u. at T is obtained through oscillatory motion of the packet. The packet is distorted in shape(not shown) while approximately retaining its original variance during the evolution. During the optimization process the total integrated probability density remained at unity up to a deviation of $10^{-5}$. The iteration algorithm takes 10 steps to achieve the results shown here at 2 CPU mins. on an IRIX Silicon Graphics Machine(Release 6.1). Within numerical precision the results were the same as obtained by solving the original Schr\"{o}dinger equation. 

\section*{5. Conclusion}

\begin{figure}[!h]
\epsfxsize=6.0in
\hspace*{0.0cm}\epsffile{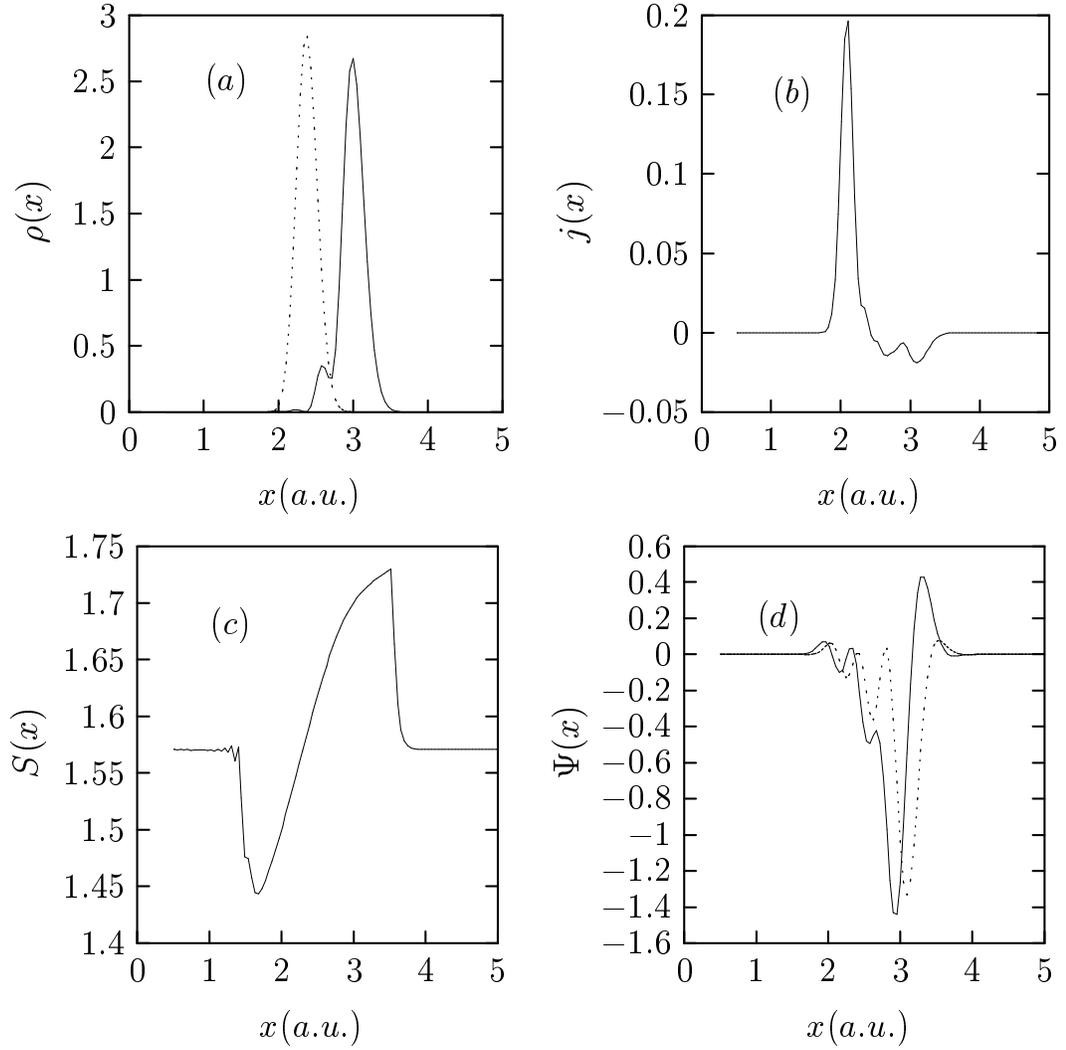}
\vspace*{-6.0cm}\caption{ Fluid dynamical variables, viz., $\rho(x)$(curve (a); dotted lines for the initial density and solid lines for the final density), j(x)(curve (b)), S(x)(curve (c)) shown as a function of x corresponding to t=T. Curve (d) shows the wave function($\Psi(x)$)(solid lines for real part and dotted lines for imaginary part) as a function of x corresponding to t=T.}
\label{fig3}
\end{figure}

This paper presents a new QFD based approach for carrying out the optimal design of control fields with an illustration for the maniputation of the HCl molecule. Our previous work [28] shows the typical smooth and monotonic behaviour of the fluid dynamical variables, viz., S and v as opposed to the typical oscillations in the wave functions where the hamiltonian was time independent. In the present case where the system is driven with an optimal time-dependent external field we have calculated the spatial dependence of j, $\rho$, S and $\Psi$ at t=T shown in Fig.\ref{fig3}. The fluid dynamical variables(Fig.\ref{fig3} curves (a), (b) and (c)) used in the present method are relatively slowly varying spatial functions compared to the wave function(Fig.\ref{fig3}, curve(d)) which apparently enhances the efficiency and the numerical saving of the present approach to controlling dynamics.

Although the illustration was for one dimension the QFD technique is directly extendable to higher dimensions, and a QFD wave packet calculation in four dimension has already been performed [28]. The alternating direction method can effectively be used with QFD for high dimensions. Comparison with FFT propagation has been performed for two dimensional systems [29], showing that QFD is capable of providing a considerable increase in efficiency(i.e., by a factor of 10 or more). Regardless of the dimension, the key advantage of OCT-QFD arises from the expected smooth nature of QFD variables. A special circumstance will arise if the control ``exactly'' leads to a bound state with nodes that fully separates one spatial region from another. In practice placing a lower limit on the density of the order of the machine precision overcomes such difficulties. Future studies need to explore the full capabilities of the computational savings afforded by OCT-QFD.

\section*{Acknowledgement}
BD thanks Drs.Jair Botina and Tak-San Ho for useful discussions. The authors acknowledge support from the NSF and DOD.
\begin{center}
{\bf APPENDIX\\Two additional forms for the cost functional and associated initial/final conditions\\}
\end{center}
The forms here are presented for reference as an alternative QFD approach. They have the advantages of simplicity and of giving equations for the Lagrange multiplier in the same form as the dynamical equations. The formalism for deriving the quations is through the Euler equations corresponding to the minimization of
\begin{eqnarray*}
I=\int _V\int _{t=0}^TF(f,f_t,\nabla f,\nabla ^2 f)dt dV \hspace*{9.0cm} (A.1)
\end{eqnarray*}
Here V denotes the volume in coordinate space.
The corresponding Euler equations and conditions on time and space are
\begin{eqnarray*}
\frac{\partial F}{\partial f}-\frac{\partial (\frac{\partial F}{\partial f_t})}{\partial t}-\nabla \cdot (\frac{\partial F}{\partial \nabla f})+\nabla ^2(\frac{\partial F}{\partial \nabla ^2f})=0 \hspace*{7.0cm} (A.2)
\end{eqnarray*}
Initial condition: $f({\bf x},0)=f_0({\bf x})$; Final condition: $(\frac{\partial F}{\partial f_t})|_{t=T}=f_T({\bf x})$ \hspace*{2.4cm} (A.3)\\
Boundary conditions on dV:\\
$f({\bf x},t)=f_B({\bf x},t)$ or ${\bf n}\cdot [\frac{\partial F}{\partial \nabla f}-\nabla \cdot (\frac{\partial F}{\partial \nabla f})]=0$; \hspace*{6.4cm} (A.4)\\
${\bf n}\cdot \nabla (f)=g_B({\bf x},t)$ or $\frac{\partial F}{\partial \nabla ^2 f}=0$ \hspace*{9.3cm} (A.5)\\
Starting with the continuity and energy conservation equations given in Eqs.(2) and (3) in the text, we rewrite them as
\begin{eqnarray*}
A_t+\nabla A \cdot \frac{\nabla S}{m}+\frac{A\nabla ^2S}{2m}=0 \hspace*{9.8cm} (A.6)
\end{eqnarray*}
\begin{eqnarray*}
AS_t+A\frac{\nabla S\cdot \nabla S}{2m}+V A-\frac{\hbar ^2}{2m}\nabla ^2A=0 \hspace*{7.8cm} (A.7)
\end{eqnarray*}
The use of the dynamical equations above in the cost functional in Eq.(8) becomes
\begin{eqnarray*}
J&=&\frac{1}{2}\omega _x(\Theta _T-\Theta ^d)^2+\omega _e\int _{t=0}^TE^2(t)dt\\\nonumber
&&-\int _V\int _{t=0}^T[[\lambda _1(A_t+\frac{\nabla A\cdot \nabla S}{m}+A\nabla ^2 S)/2m] \hspace*{6.4cm} (A.8)\\\nonumber 
&&+[\lambda _2(AS_t+A\frac{\nabla S \cdot \nabla S}{2m}+ VA-\mu E(t)A -\frac{\hbar ^2}{2m}\nabla ^2 A)]]dt dV
\end{eqnarray*}
where $\Theta _T=\int _V\Theta ({\bf x})A^2({\bf x},T)dV$. The corresponding Euler equations are obtained from the formulas in $(A2)$ for arbitrary variations of A, S, $\lambda _1$, $\lambda _2$ and $E(t)$ as
\begin{eqnarray*}
A_t+{\bf v}\cdot \nabla A=-A\nabla \cdot {\bf v}/2
\end{eqnarray*}
\begin{eqnarray*}
S_t+{\bf v}\cdot \nabla S/2=-V +\mu E(t)+\frac{\hbar ^2}{2m}\nabla ^2A/A
\end{eqnarray*}
\begin{eqnarray*}
\lambda _{1t}+{\bf v}\cdot \nabla \lambda _1=-[\lambda _1\frac{\nabla.{\bf v}}{2}+\frac{\hbar ^2}{2m}\lambda _2(\nabla ^2A/A-\nabla ^2\lambda _2)/\lambda _2)] \hspace*{4.0cm} (A.9)
\end{eqnarray*}
\begin{eqnarray*}
\lambda _{2t}+{\bf v}\cdot \nabla \lambda _2=[\lambda _2\frac{\nabla.{\bf v}}{2}+\frac{1}{2m}\lambda _1(\nabla ^2A/A-\nabla ^2\lambda _1)/\lambda _1)]
\end{eqnarray*}
\begin{eqnarray*}
\omega _eE(t)+\int _V\lambda _2\mu A dV=0
\end{eqnarray*}
Following the formulas given in Eq.(A3) to (A5), the corresponding initial and final conditions become\\

$A({\bf x},0)=A_0({\bf x})$; $S({\bf x},0)=S_0({\bf x})$;\\

$\lambda _1({\bf x},T)+2\omega _x(\Theta _T-\Theta ^d)A({\bf x},T)\Theta ({\bf x})$; $\lambda _2({\bf x},T)=0$ \hspace*{4.0cm} (A.10)\\

The first two formulas in A.9 are equivalent to the Schr\"{o}dinger equation. They can be transformed into various QFD forms in terms of $\rho $, {\bf v} and {\bf j} as in Eqs.(2) to (5) in the main text. The third and fourth equations in A.9 are the basic equations for the Lagrange multiplier functions. They are in the same flux conservation form as the QFD equations. Indeed, the third equation multiplied by $\lambda _1$ can be rearranged in the form of mass conservation for $\Lambda _1=\lambda _1^2$ as
\begin{eqnarray*}
\Lambda _{1t}+\nabla \cdot (\Lambda _1{\bf v})=-[\frac{\hbar ^2}{m}\lambda _1\lambda _2(\nabla ^2A/A-\nabla ^2\lambda _2/\lambda _2)]
\end{eqnarray*}
The above derivation also can be obtained starting with the usual Schr\"{o}dinger equation and its complex conjugate. Following this approach the cost functional below assures that the external field is real
\begin{eqnarray*}
J&=&\frac{1}{2}\omega _x(\Theta _T-\Theta ^d)^2+\omega _e\int _{t=0}^TE^2(t)dt\\\nonumber
&&-\int _V\int _{t=0}^T[\lambda ^*[i\Psi _t+\frac{\hbar ^2}{2m}\nabla ^2 \Psi -V\Psi -\mu E(t)\Psi]\\\nonumber
&&+\lambda [-i\Psi ^*_t+\frac{\hbar ^2}{2m}\nabla ^2\Psi ^*-V\Psi ^*-\mu E(t)\Psi ^*]]dt dV
\end{eqnarray*}
With the substitution $\Psi = Aexp(iS)$, the cost functional reduces to the one in Eq.(A.8) with $\lambda = \lambda _1 +i\lambda _2$.

\section*{References}
\begin{enumerate}
\item S. A. Rice, Science, {\bf 258}, 412 (1992)
\item D. J. Tannor and S. A. Rice, J. Chem. Phys. {\bf 83}, 5013 (1985)
\item A. P. Peire, M. A. Dahleh and H. Rabitz, Phys. Rev. {\bf A 37}, 4950 (1988)
\item D. J. Tannor, R. Kosloff and S. A. Rice, J. Chem. Phys., {\bf 85}, 5805 (1986)
\item R. Demiralp and H. Rabitz, Phys. Rev. {\bf A 47}, 809 (1993)
\item J. Botina and H. Rabitz, J. Chem. Phys. {\bf 104}, 4031 (1996)
\item S. Shi and H. Rabitz, Comput. Phys. Comm., {\bf 63}, 71 (1991)
\item S. Shi and H. Rabitz, Chem. Phys., {\bf 139}, 185 (1989)
\item W. Zhu, J. Botina and H. Rabitz, J. Chem. Phys., {\bf 108}, 1953 (1998)
\item Y. Ohtsuki, H. Kono and Y. Fujimura, J. chem. Phys., {\bf 109}, 9318 (1998)
\item J. Cao and K. R. Wilson, J. Chem. Phys., {\bf 107}, 1441 (1997)
\item D. J. Tannor and S. A. Rice, Adv. Chem. Phys., {\bf 70}, 441 (1988)
\item R. Kosloff, S. A. Rice, P. Gaspard, S. Tersigni and D. J. Tannor, Chem. Phys., {\bf 139}, 201 (1989)
\item P. Brumer and M. Shapiro, Faraday Discuss. Chem. Soc., {\bf 82}, 177 (1986)
\item P. Brumer and M. Shapiro, Annu. Rev. Phys. Chem., {\bf 43}, 257 (1992)
\item T. Baumert and G. Gerber, Isr. J. Chem., {\bf 34}, 103 (1994)
\item H. Rabitz and S. Shi in {\bf Advances in Molecular Vibration and Collisional  Dynamics}, Vol. {\bf 1A}, 187 (1991)
\item Judson and H. Rabitz Phys. Rev. Lett. {\bf 68}, 1500 (1992)
\item A. Assion, T. Baumert, M. Bergt, T. Brixner, B. Kiefer, V. Strehle and G. Gerber, Science, {\bf 282}, 919 (1998)
\item C. J. Bardeen, V. V. Yakovlev, K. R. Wilson, S. D. Carpenter, P. M. Weber and W. S. Warren, Chem. Phys. Lett., {\bf 280}, 151 (1997)
\item A. Assion, T. Baumert, V. Seyfried and G. Gerber in {\bf Ultrafast Phenomena} edited by J. Fujimoto, W. Zinth, P. F. Barbara and W. H. Knox, Springer, Berlin (1996)
\item J. L. Herek, A. Materny and A. H. Zewail, Chem. Phys. Lett., {\bf 228}, 15 (1994)
\item V. D. Kleiman, L. Zhu, J. Allen and R. J. Gordon, J. Chem. Phys., {\bf 103}, 10800 (1995)
\item G. Q. Xing, X. B. Wang, X. Huang anf R. Bersohn, J. Chem. Phys., {\bf 104}, 826 (1996)
\item A. Shnitman, I. Sofer, I. Golub, A. Yogev, M. Shapiro, Z. Chen and P. Brumer, Phys. Rev. Lett., {\bf 76}, 2886 (1996)
\item (a) D. Bohm, Phys. Rev., {\bf 85}, 166 (1952)\\
      (b) D. Bohm, Phys. Rev., {\bf 85}, 180 (1952)
\item D. Bohm, B. J. Hiley and P. N. Kaloyerou, Phys. Rep., {\bf 144}, 321 (1987)
\item Bijoy K. Dey, Attila Askar and H. Rabitz, J. Chem. Phys., {\bf 109}, 8770 (1998)
\item F. S. Mayor, A. Askar and H. Rabitz, J. Chem. Phys., {\bf 3}, 2423 (1999)
\item R. L\"{o}hner, K. Morgan, J. Peraire, M. Vahdari, Int. J. Num. Methods Fluid, {\bf 7}, 1093 (1987)
\item G. G. Balint-Kurti, C. L. Ward and C. C. Marston, Comput. Phys. Comm., {\bf 67}, 285 (1991)
\item C. C. Marston and G. G. Balint-Kurti, J. Chem. Phys., {\bf 91}, 3571 (1989)
\item J. F. Olgilvie, Proc. R. Soc. Lond., {\bf A 378}, 287 (1981)
\item G. G. Balint-Kurti, R. N. Dixon and C. C. Marston, J. Chem. Soc. Faraday Trans., {\bf 86}, 1741 (1990)
\item J. P. Boris and D. L. Book, Methods in Comp. Phys., {\bf 16}, 85 (1976)
\item W. H. Press, B. P. Flannery, S. A. Teukolsky and W. T. Vetterling, {\bf Numerical Recipes}, Cambridge University, New York, (1992)
\end{enumerate}

\end{document}